\documentclass[%
reprint,
 amsmath,amssymb,
 aps,pra,
]{revtex4-1}
\usepackage{graphicx}
\usepackage{dcolumn}
\usepackage{bm}

\usepackage{physics}
\usepackage{gensymb}
\usepackage{subfigure}
\usepackage{flushend}
\usepackage{nicefrac}
\usepackage{siunitx}
\usepackage{array}
\newcolumntype{P}[1]{>{\centering\arraybackslash}p{#1}}
\begin{document}

\linespread{1}
\preprint{APS/123-QED}

\title{Verifying Thermodymanic Equilibrium of Molecular Manifolds: Kennard-Stepanov Spectroscopy of a Molecular Gas}

\author{Stavros Christopoulos$^{1,2}$}
\author{Dominik M{\"o}ller$^{1}$}%
\author{Roberto Cota$^{1}$}
\author{Benedikt Gerwers$^{1}$}
\author{Martin Weitz$^{1}$}
\affiliation{$^{1}$Institut f{\"u}r Angewandte Physik der Universit{\"a}t Bonn, Wegelerstr. 8, D-53115 Bonn, Germany\\$^{2}$Department of Science, American University of the Middle East, Egaila, Kuwait
}%



\date{\today}

\begin{abstract}

The degree of thermalization of electronically excited state manifolds of an absorber can be tested via optical spectroscopy. In the thermalized manifolds case, the ratio of absorption and emission is expected to follow a universal Boltzmann-type frequency scaling, known as the Kennard-Stepanov relation. Here we investigate absorption and emission spectral profiles of rubidium, caesium and potassium molecular dimers in high pressure argon buffer gas environment and study the effect of collisionally induced redistribution. We find that, despite the use of nonlinear excitation techniques, the ratio of absorption and emission well follows the Kennard-Stepanov scaling for a variety of molecular transitions. We conclude that the upper electronic state rovibrational manifold of the molecular gas is well in thermodynamic equilibrium. Further, we demonstrate an accurate, calibration-free determination of the gas temperature from the measured spectroscopic data.
\end{abstract}


\maketitle



\section{INTRODUCTION}

The spectroscopic study of molecular gases is a field of great significance in physics and chemistry, revealing a wide range of properties of these building blocks of matter \cite{MPbook}. Optical spectroscopy of molecular vapor usually operates far from thermal equilibrium conditions, as is most evident from the vibrational wavepacket dynamics observed in ultrafast measurements, whose evolution do not approach an equilibrium distribution in the Born-Oppenheimer potentials throughout the timespan set by the upper electronic state spontaneous decay \cite{Zewail,FFbook}. Optical excitation here creates a non-equilibrium rovibrational distribution within the electronically excited state manifold and the Franck-Condon factors of the instantaneous distribution determine the observed spectroscopic signal. Interactions of an absorber with the environment can lead to relaxations, and for the case of atomic vapors it is known that collisions with rare gas atoms are extremely elastic \cite{Speller}.

In this Letter we examine the applicability of the Kennard-Stepanov relation, a Boltzman-type frequency scaling of the ratio of absorption and emission amplitudes \cite{Kennard,Stepanov}, in molecular alkali dimers at high-pressure noble buffer gas conditions. It has been pointed out that this thermodynamic relation is connected to both the sublevel manifolds of the ground and the electronically excited states being in thermal equilibrium \cite{Sawicki}. A  Kennard-Stepanov scaling has been observed for many solid and liquid phase systems, including semiconductor quantum wells \cite{Ihara} and doped glasses \cite{Sawicki}, as well as a wide variety of photoactive biomolecules \cite{Croce,Dau} and dyes \cite{Dobek,Kawski}. Our group has recently observed Kennard-Stepanov scaling of the ratio of absorption and emission in an atomic vapor at a few hundred bar of noble buffer gas pressure \cite{Moroshkin}.

We here report  measurements of absorption and emission spectra of rubidium, caesium and potassium molecular dimers at high pressure noble buffer gas conditions and two-dimensional spectroscopy in this regime. The emission spectra are recorded following non-resonant excitation, while the measured absorption spectra correspond to the usual one-photon extinction process. It is demonstrated that the Kennard-Stepanov relation is well fulfilled in the molecular-buffer gas ensembles for a variety of transitions, species and experimental parameters, providing direct evidence for a thermalized rovibrational distribution in the electronically excited state molecular manifolds. Further, nonlinear excitation has been used for the first time in Kennard-Stepanov studies, generalizing the applicability of the relation. The here observed slope of the Boltzmann-type frequency scaling is found to be in good agreement with the expected value assuming the known cell temperature. Future accurate measurements of this scaling in high-pressure buffer gas systems could serve as a novel means for frequency based determinations of temperature and the Boltzmann constant.

\section{MODEL AND THEORY}

For a simple model of the molecular system subject to frequent buffer gas collisions, consider an electronic two-level system with ground state $\ket{g}$ and  electronically excited state $\ket{e}$, each of which are subject to additional rovibrational substructure, see Fig.~\ref{fig:jec}(a). If the thermalization rate by rare gas collisions is much faster than both the laser coupling as well as spontaneous decay, the sublevels in both ground and electronically excited manifolds will be Boltzmann distributed. This yields occupation probabilities 
$p^g_i=n_g\cdot exp\left[\nicefrac{-e_i}{k_BT}\right]$ and $p^e_j=n_e\cdot exp\left[\nicefrac{-e_j}{k_BT}\right]$
for the sublevels with energy $e_i$ and $e_j$ corresponding to the lowest rovibrational level in the ground and electronically excited state respectively. Here, $n_g$ and $n_e$ are normalization factors, while $T$ denotes the temperature. Assuming such thermalized distributions of sublevels, following \cite{Sawicki,Klaers2}, the Kennard-Stepanov law can be derived. Energy conservation implies that $e_i+ h\nu= h\nu_0+e_j$, where $\nu$ and $\nu_0$ denote optical frequency and zero-phonon line of the molecule respectively, from which we obtain
$p^e_j=p^g_i\frac{n_e}{n_g}\cdot exp\left[\nicefrac{- h\left(\nu-\nu_0\right)}{k_BT}\right]$. Further, let $B_{ij}$ and $A_{ji}$ denote the Einstein coefficients for absorption and spontaneous emission respectively between individual rovibrational sublevels. The ratio between the frequency dependent absorption $\alpha(\nu)$ and emission $f(\nu)$ coefficients is:

\begin{equation}\label{eq:0}
\frac{f(\nu)}{\alpha(\nu)} \propto \frac{A(\nu)}{B(\nu)} = \frac{\sum_{i,j}p_jA_{ji}}{\sum_{i,j}p_iB_{ij}} \propto e^{\left(\frac{-h\nu}{k_BT}\right)},
\end{equation} 
where in the last step we have applied the Einstein A-B relation, which here takes the form $A_{ji}=B_{ij}$, and the above relation between the occupation probabilities $p^e_i$ and $p^g_j$. This verifies the Boltzmann-type Kennard-Stepanov relation between absorption and emission. We will below mostly use the Kennard-Stepanov law in its logarithmic form
\begin{equation} \label{eq:1}
\ln\bigg[\frac{\alpha (\nu)}{f(\nu)}\bigg] = \frac{h}{k_{B}T}\nu + C
\end{equation}
where $C=C(T)$ is a temperature dependent constant. 

\begin{figure}[t!]
\centering
\includegraphics[scale=0.256]{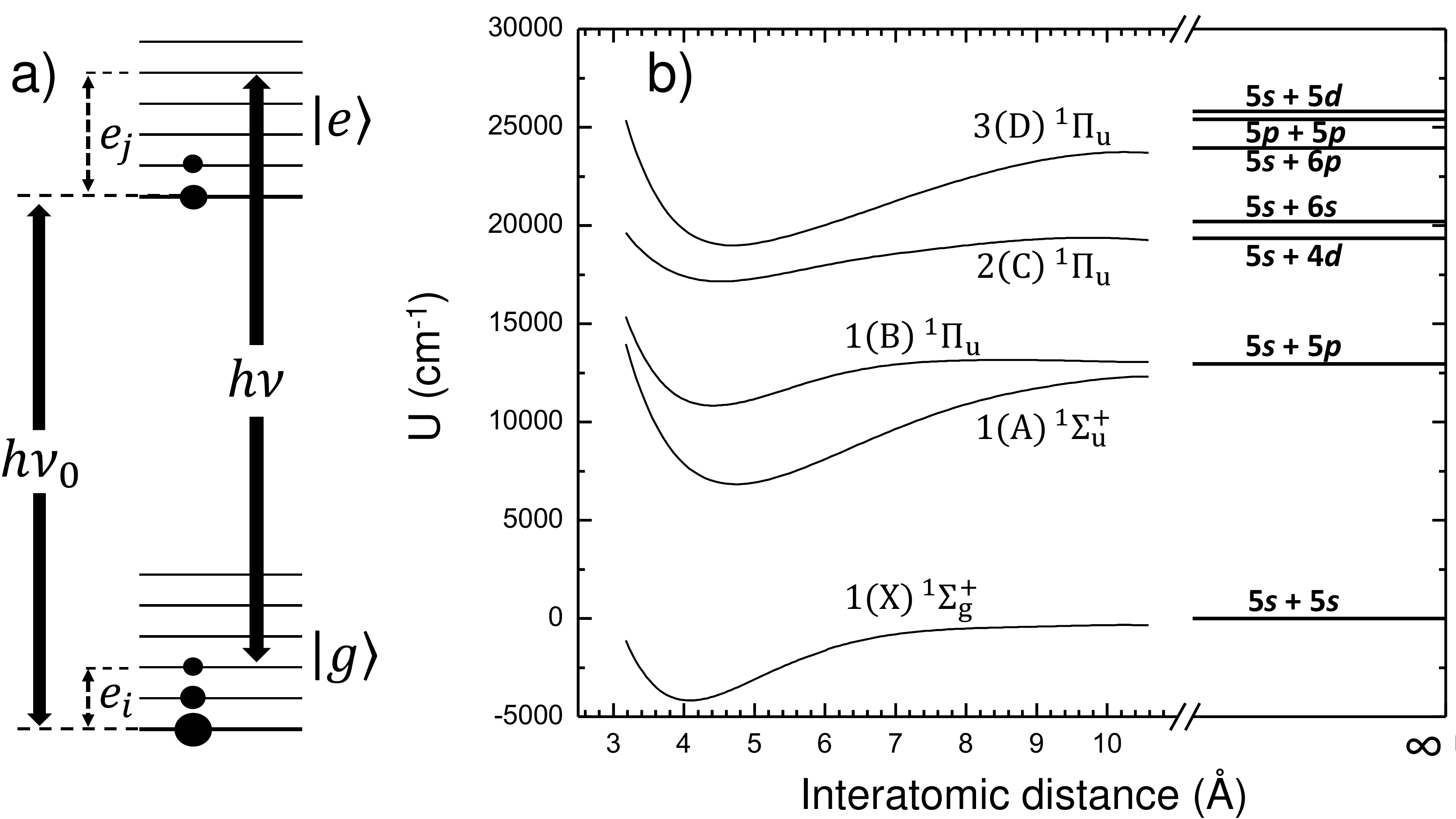}\label{fig:jec}
\caption{\label{fig:jec} (a) Simplified Jablonski diagram of a molecule with an electronic ground state and an electronically excited state, each of which is subject to additional rovibrational substructure. The size of the dots is to indicate the occupation of sublevels in the presence of optical excitation and thermalization by buffer gas collisions. (b) Calculated energy versus interatomic distance in the rubidium dimer system \cite{Spiegelmann}.}
\end{figure}

For a system with thermalized rovibrational distributions in both upper and lower electronic states, we thus expect a linear scaling of the logarithm versus optical frequency, with the corresponding slope giving direct access to the temperature measured in frequency units. While atomic systems have been investigated in high pressure buffer gas conditions up to a few hundred bars \cite{Vogl2}, the study of molecular systems under these conditions is basically unknown. Figure ~\ref{fig:jec}(b) shows calculated  energy levels of the rubidium dimer system versus the distance between the rubidium nuclei, as predicted in the absence of a buffer gas \cite{Spiegelmann}. Here, the ground and electronically excited states potential curves do not have their minima at the same internuclear distance, so that, in the absence of thermalization, optical excitation would create a highly non-equilibrium vibrational distribution in the upper electronic state. In the dense buffer gas environment of our system, frequent collisions between rubidium dimers and rare gas atoms are expected to cause relaxations. Our experiment probes the degree of thermalization of vibrational sublevels in ground and electronically excited states within such molecular potential curves.

\section{EXPERIMENTAL RESULTS AND DISCUSSION}

Our experimental apparatus uses high-pressure, stainless steel cells, where optical access is achieved by sapphire optical windows, similarly as in \cite{Moroshkin,Vogl}. An ampoule containing 1 g of either rubidium, caesium or potassium metal is placed into the cell reservoir and subsequently the cell is filled with high-pressure argon gas. Heating belts, wrapped around both cell and its reservoir, allow us to reach temperatures of up to 730 K, as necessary to obtain sufficiently high vapor density of molecules in the pressure-broadened system \cite{Vdovic}. In this temperature range, the vapor pressure ratio of monomers to dimers increases exponentially with temperature and reaches values of up to $\SI{3}{\percent}$. The vapor density of rubidium dimers at 673 K is $3 \cdot 10^{15}$ $\mathrm{cm^{-3}}$, for caesium dimers at 600 K it is $6 \cdot 10^{14}$ $\mathrm{cm^{-3}}$, while for potassium dimers at 723 K it is $2 \cdot 10^{15}$ $\mathrm{cm^{-3}}$. Thermocouples are used to measure the cell temperature, which is controlled to within $\SI{\pm0.5}{K}$. Emission spectroscopy of molecular levels is performed upon non-resonantly exciting the pressure-broadened ensemble with radiation from a cw Ti:sapphire laser, tunable in a wavelength range between $\SI{770}{\nano\meter}$ and $\SI{850}{\nano\meter}$. We typically use laser powers of $0.1 - \SI{1.5}{\watt}$ on a 1 - 2 mm beam diameter. These intensities are sufficiently low to avoid significant collisionally-induced, redistribution laser cooling or heating effects \cite{Vogl}. The fluorescence is collected in a direction orthogonal to the excitation beam and focused onto a multimode fibre coupled to a spectrometer.

Measurements of the molecular absorption spectrum of the pressure-broadened ensemble are conducted by alternatively using radiation of a blue spectral range light emitting diode as well as a broadband halogen lamp, which allows for resonant excitation of molecular bands in different wavelength ranges. The corresponding sample absorption is determined by recording reference and transmission spectra collected in optical propagation direction. We note that the optical densities of the molecular lines in the buffer gas system are moderate (typically OD $<$ 1), so that reabsorption effects are small.

\begin{figure}[t!]
\centering
\includegraphics[scale=0.39]{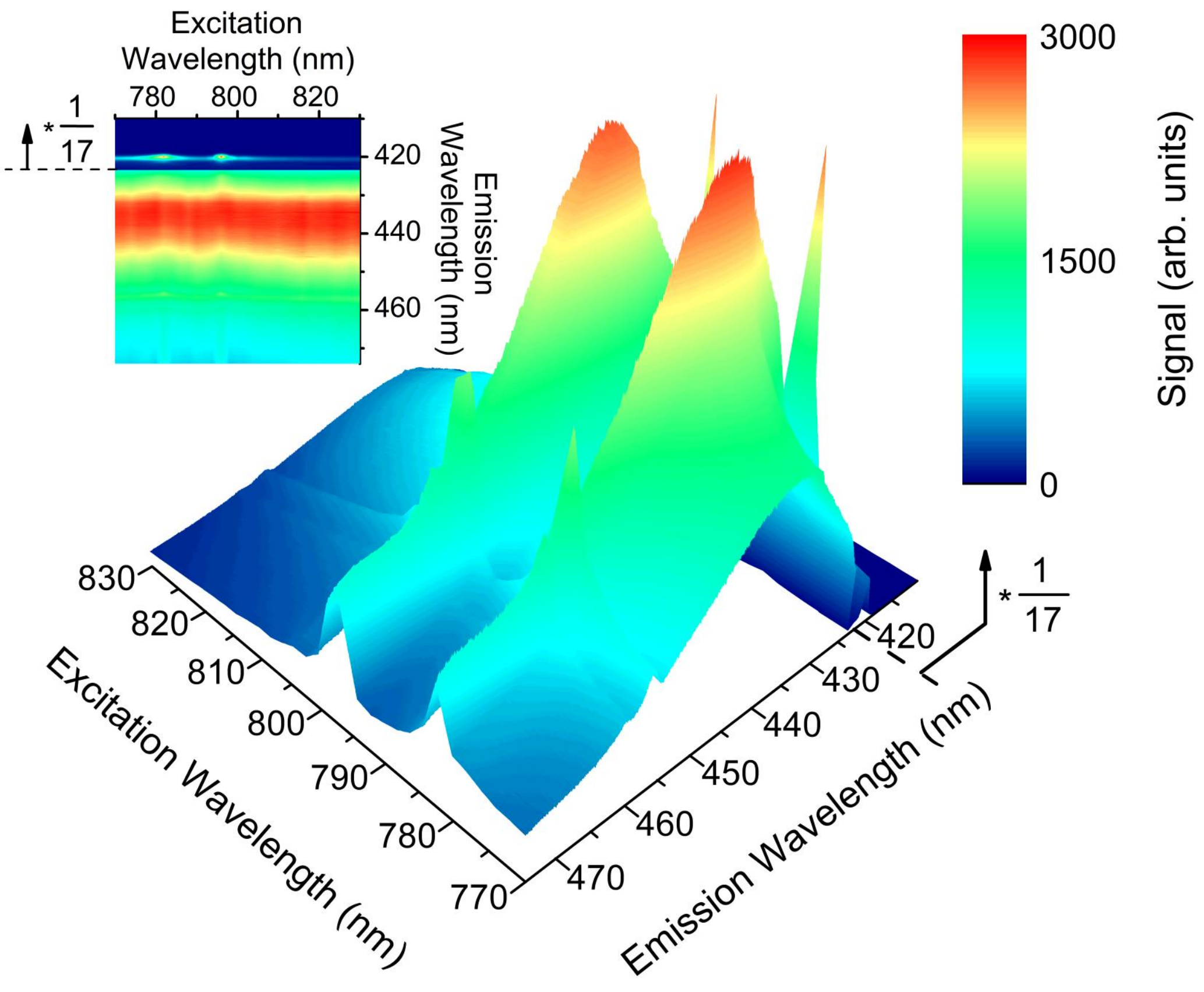}
\caption{(color online). Measured emission spectra profile for the rubidium system near the $1(\mathrm{X}){}^1\Sigma_{\mathrm{g}}^{+} - 3(\mathrm{D}){}^1\Pi_{\mathrm{u}}$ dimer transition versus excitation wavelength. The detected spectral intensity is encoded in the color scale. For emission wavelengths below 423 nm, corresponding near to the atomic rubidium second principal series, the intensity is demagnified by a factor of 17. Inset: Normalized emission spectral profile versus excitation wavelength, using the same color scale and demagnification factor below 423 nm.}
\label{fig:ed}
\end{figure}

To begin with, we investigate the spectral redistribution in the pressure-broadened ensemble by recording molecular emission spectra for different excitation wavelengths. Figure \ref{fig:ed} (main plot) shows experimental results for such a two-dimensional spectroscopic measurement for the case of the rubidium dimer $1(\mathrm{X}){}^1\Sigma_{\mathrm{g}}^{+} - 3(\mathrm{D}){}^1\Pi_{\mathrm{u}}$ transition at $\SI{170}{\bar}$ argon buffer gas pressure, giving the variation of the corresponding emission spectrum versus the excitation wavelength. All given values for wavelengths in this work are vacuum wavelengths. The excitation laser here is tuned over a broad wavelength range including the atomic rubidium $D_1$ and $D_2$ lines which appear at the pressure shifted wavelengths near $\SI{796}{\nano\meter}$ and $\SI{781}{\nano\meter}$ respectively. We observe that when exciting resonantly to the two clearly visible atomic $D$-lines, both the $1(\mathrm{X}){}^1\Sigma_{\mathrm{g}}^{+} - 3(\mathrm{D}){}^1\Pi_{\mathrm{u}}$ molecular transition, as well the atomic Rb(6$p$)$\rightarrow$Rb(5$s$) doublet, exhibit enhanced fluorescence. The inset  (Figure \ref{fig:ed} - left hand side) gives the data in a normalized form, revealing that the molecular emission spectra, with their emission maximum near 435 nm, have nearly the same form independently of the excitation wavelength. This is a typical signature for redistribution within the electronically excited state manifold to a thermalized distribution \cite{Kasha}. A residual broadening of the spectra and a shift towards longer wavelengths is visible for the case of excitation resonant to the atomic $D$-lines. Further measurements on the rubidium system are therefore conducted using excitation wavelengths far detuned from the atomic resonances.

Previous experiments with alkali vapors carried out at moderate noble buffer gas pressures have shown that excitation at the $D$-lines can efficiently populate high-lying molecular states via collision induced energy pooling effects \cite{Allegrini3,Zhou,He}. These include energy transfer from an excited atom to a ground state molecule, as well as collisional molecule formation by one excited and one ground state atom. We conjecture that the here observed signal of $\mathrm{Rb}_2$($3(\mathrm{D}){}^1\Pi_{\mathrm{u}}$) molecules in the high pressure buffer gas system is also due to such formation channels. Note that the asymptote of this molecular level is energetically close to several different combinations of excited atom pairs, see Fig. 1(b). In the high pressure system also far detuned radiation can lead to the excitation of Rb(5p) atoms due to collisional redistribution \cite{Vogl,Moroshkin}. Upon the collision of two Rb(5p) atoms, energy pooling is expected to populate the Rb(5d) and Rb(6p) states \cite{Barbier,Ban}. The molecular $\mathrm{Rb}_2$($3(\mathrm{D}){}^1\Pi_{\mathrm{u}}$) state can then be populated through the channels: \hfill \break


(i)Rb(6p,5d)+$\mathrm{Rb_2{}}(1(\mathrm{X}){}^1\Sigma_{\mathrm{g}}^{+})\rightarrow\mathrm{Rb_{2}}(3(\mathrm{D}){}^1\Pi_{\mathrm{u}})+\mathrm{\Delta E_{1}}$,\hfill \break

(ii)Rb(6p,5d)+Rb(5s)$\rightarrow\mathrm{Rb_{2}}(3(\mathrm{D}){}^1\Pi_{\mathrm{u}})+\mathrm{\Delta E_{2}}$,\hfill \break

\noindent where $\mathrm{\Delta E_{1,2}}$ are the corresponding energy defects. The latter process can be also facilitated by an argon atom in a three-body collision scheme. Our conjecture of the molecular $\mathrm{Rb}_2$($3(\mathrm{D}){}^1\Pi_{\mathrm{u}}$) level population being dominated by atomic excitation channels is supported by the observation that the measured molecular fluorescence is enhanced near the atomic $D$-lines (Fig.\ref{fig:ed}). We note however that the nature of the redistribution is such that the produced final state is independent from the excitation channel.


\begin{figure}[b!]
\centering
\includegraphics[scale=0.26]{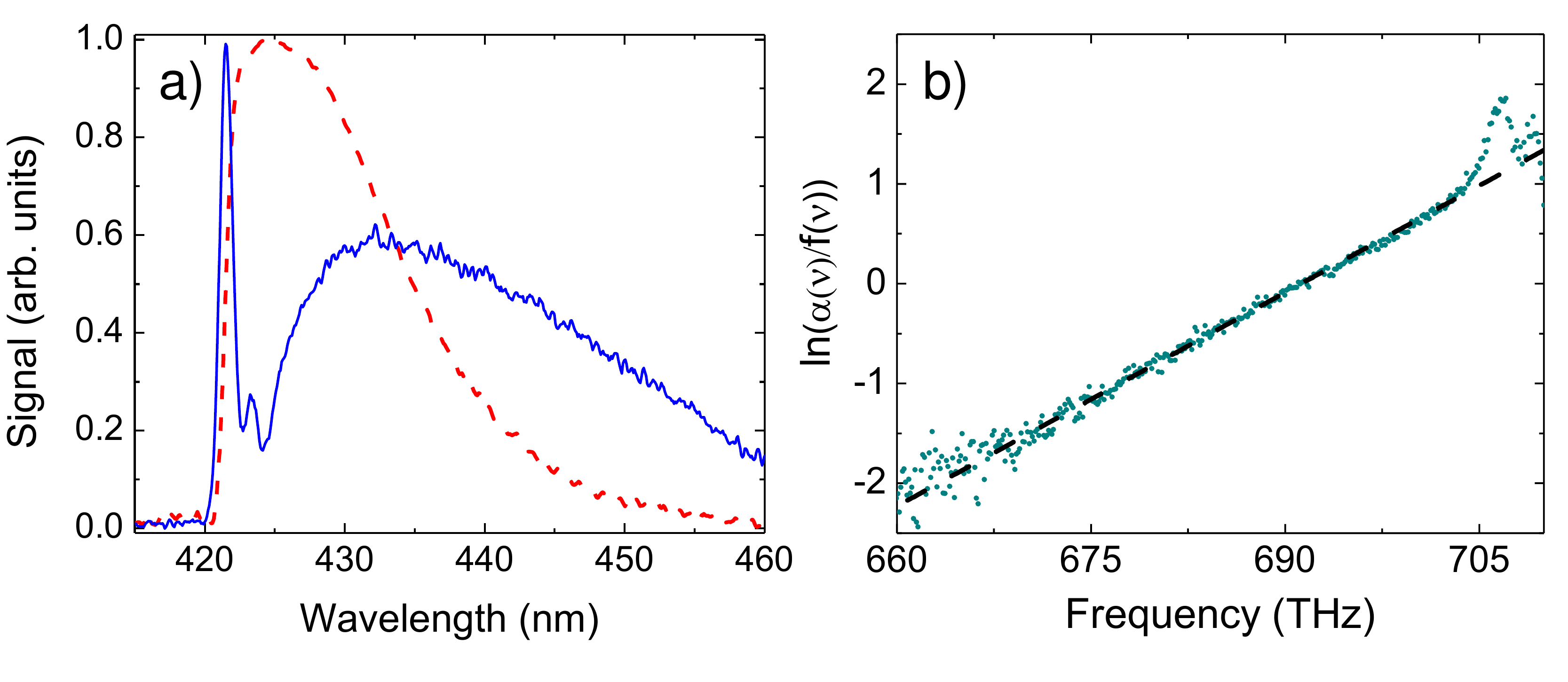}
\caption{(color online). (a) Rubidium dimer emission (solid blue line) and absorption (dashed red line) spectra for cell temperature of 673 K and 135 bar of argon buffer gas pressure. To record the emission spectrum the sample was irradiated with a laser beam ($I=1.27$ $\mathrm{W/mm^{2}}$) near $\SI{800}{\nano\meter}$ wavelength, which is far detuned from the atomic $D$-line resonances. (b) Logarithmic ratio of the measured absorption and fluorescence profiles versus frequency (green dots), along with a linear fit (dashed black line).}
\label{fig:faks1}
\end{figure}

We initially test for the applicability of the Kennard-Stepanov relation for the case of the $1(\mathrm{X}){}^1\Sigma_{\mathrm{g}}^{+} - 3(\mathrm{D}){}^1\Pi_{\mathrm{u}}$ rubidium molecular dimer transition. The solid blue line in Fig. \ref{fig:faks1}(a) is an experimental emission spectrum for laser irradiation far detuned from the atomic $D$-lines, while the dashed red line is an absorption spectrum recorded using a broadband source, both at $\SI{135}{\bar}$ argon buffer gas pressure and $\SI{673}{\kelvin}$ cell temperature. In the fluorescence spectrum, besides the broad molecular transition centered at around $\SI{435}{\nano\meter}$, the sharper atomic Rb($6p$) $\rightarrow$ Rb($5s$) doublet near $\SI{421.4}{\nano\meter}$ and $\SI{423.1}{\nano\meter}$ respectively is also resolved, spectrally overlapping with the molecular line. We conduct power dependence measurements to study the relative contributions of atomic and molecular lines to the total emission and find that high intensities favor the emission originating from the atomic lines, which exhibit non-linear behavior. As discussed above, the spectral shape of the molecular emission is observed to be, in good approximation, independent from excitation parameters.

Figure \ref{fig:faks1}(b) depicts the logarithm of the ratio of measured absorption $\alpha(\nu)$ and emission $f(\nu)$ versus the optical frequency. A clear linear behavior is found for the ratio within the optical frequency range between $\SI{660}{\tera\hertz}$ and $\SI{704}{\tera\hertz}$, see also the (linear) fit shown by the black dashed line. This verifies the Kennard-Stepanov law for this transition and is interpreted as one line of evidence for the formation of a thermalized distribution in the rovibrational sublevels in both lower and upper electronic molecular states. The visible deviation on the lower frequency side of the presented range is regarded as instrumental, caused by the here rapidly decreasing absorption signal at large detunings. On the other hand, on the high frequency side, where the atomic Rb($6p$) level is populated, a deviation from the Kennard-Stepanov scaling is also observed. This could be attributed to finite quantum efficiency for the, compared to the $D$-lines transitions, higher excited upper electronic state within the high pressure buffer gas environment. 
The fitted slope $\nicefrac{h}{k_BT}$, see Eq. \ref{eq:1}, of the experimental data of Fig. \ref{fig:faks1}(b) allows us to extract the spectral temperature, which is found to be $T_{extr}=\SI{672.6 \pm 3.3}{\kelvin}$. This value is in very good agreement with the above noted measured cell temperature.

\begin{figure}[t!]
\centering
\includegraphics[scale=0.26]{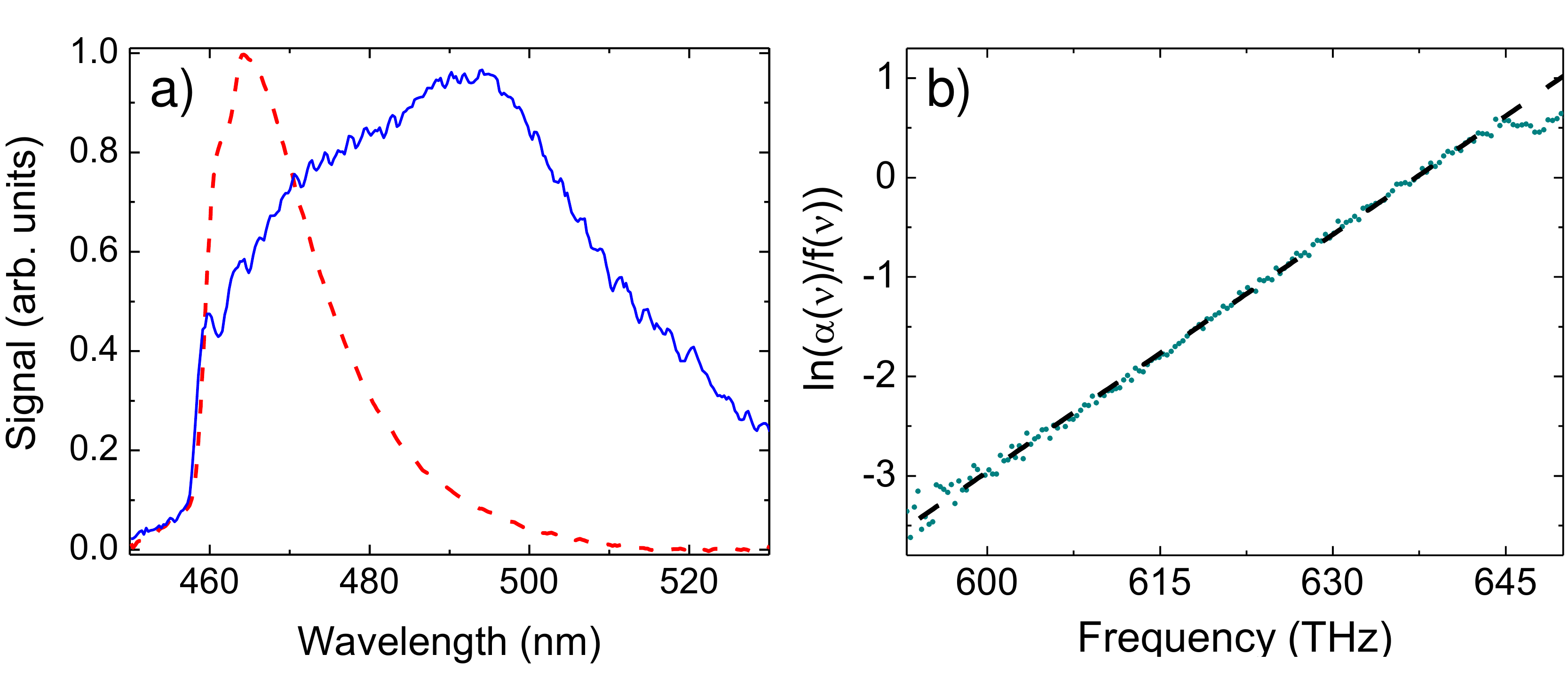}
\caption{(color online). (a) Caesium dimer emission (solid blue line) and absorption (dashed red line) spectra for $\SI{603}{\kelvin}$ cell temperature and $\SI{240}{\bar}$ argon buffer gas pressure. (b) Logarithm of the ratio of the measured absorption and fluorescence profiles versus frequency (green dots), along with a linear fit (dashed black line).}
\label{fig:faks2}
\end{figure}

We next investigate Kennard-Stepanov scaling for the case of the $1(\mathrm{X}){}^1\Sigma_{\mathrm{g}}^{+} - 3(\mathrm{E}){}^1\Sigma_{\mathrm{u}}^{+}$ transition of the caesium dimer. Corresponding emission (solid blue line) and absorption (dashed red line) spectra are shown in Fig. \ref{fig:faks2}(a) for $T=\SI{603}{\kelvin}$ and $\SI{240}{\bar}$ argon buffer gas pressure. For this measurement, a lower excitation intensity ($I=0.42$ $\mathrm{W/mm^{2}}$) has been used to record the emission spectra than in the case of Fig. \ref{fig:faks1}(a), leading to the second principal series atomic doublet (Cs($6s$) $\rightarrow$ Cs($7p$)) near $\SI{459.7}{\nano\meter}$ and $\SI{463.8}{\nano\meter}$ being barely visible. For this data set, we have correspondingly been able to use higher buffer gas pressure than in the case shown in Fig. \ref{fig:faks1}, to further extend the wavelength range of spectral overlap between absorption and emission data. Figure \ref{fig:faks2}(b) shows the logarithm of the ratio of absorption and emission versus frequency for the caesium data, in which we observe a linear behavior within an optical frequency range of  more than $4k_BT/h$ within the expectations for a Kennard-Stepanov scaling. The fitted inverse slope yields here a spectral temperature of $T_{extr}=\SI{603.1\pm 2.5}{\kelvin}$, a value in very good agreement with the measured cell temperature.

\renewcommand{\arraystretch}{1.4}
\begin{table}[b!]
\centering 
\begin{tabular}{ |P{1.3cm}||P{1.5cm}|P{1.3cm}|P{1.4cm}||P{1.65cm}|  }
 \hline
 Molecule & Transition & $T_{amb}$ (K) & $T_{extr}$ (K) & Range\\
 \hline
$\mathrm{Rb_{2}}$   &$\mathrm{X} - \mathrm{D}$    & 673 &   $\SI{672.6\pm 3.3}{} $ & $2.7k_{B}T$\\
$\mathrm{Rb_{2}}$   &$\mathrm{X} - \mathrm{D}$    & 623 &   $\SI{624.5\pm2.0}{} $ & $ 2.2k_{B}T$\\
$\mathrm{Rb_{2}}$   &$\mathrm{X} - \mathrm{D}$    & 593 &   $\SI{590.9\pm5.9}{} $ & $ 1.2k_{B}T$\\
$\mathrm{Rb_{2}}$   &$\mathrm{X} - \mathrm{A}$    & 583 &   $\SI{583.2\pm2.9}{} $ & $1.9k_{B}T$\\
$\mathrm{Cs_{2}}$   &$\mathrm{X} - \mathrm{E}$   & 603 &   $\SI{603.1\pm2.5}{} $ & $ 4.1k_{B}T$\\
$\mathrm{K_{2}}$   &$\mathrm{X} - \mathrm{C}$   & 723 &   $\SI{724.5\pm6.9}{} $ & $ 2.4k_{B}T$\\
 \hline
\end{tabular}
\caption{Aggregated results for the temperature obtained from the inverse slope of the logarithmic ratio of  measured absorption and emission ($T_{extr}$) along with the corresponding measured cell temperatures ($T_{amb}$). Argon buffer gas pressures of 130 - 240 bar are used for the different data sets. The right column shows the frequency range in energy units, within which Kennard-Stepanov scaling is observed.}
\label{tab:table}
\end{table}

Table \ref{tab:table} summarizes our results for the spectral temperature derived from absorption and emission spectra of the pressure-broadened ensemble for different molecular transitions. Further, different cell temperatures have been investigated. Within experimental uncertainties, the spectrally extracted temperature values in all cases are in good agreement with the quoted, measured cell temperatures, which here verifies the Boltzmann-type Kennard-Stepanov scaling in the molecular high-pressure buffer gas system. We regard the observation of this scaling for different lines, three different molecular species, as well as different cell temperatures as strong evidence that both molecular ground and electronically excited vibrational manifolds are in thermal equilibrium.

Alternatively to extracting the temperature, one may, from the measured absorption and emission data also determine the Boltzmann-constant, when using the measured cell temperature in Eq. \ref{eq:1}. From e.g. our data for the caesium dimer $1(\mathrm{X}){}^1\Sigma_{\mathrm{g}}^{+} - 3(\mathrm{E}){}^1\Sigma_{\mathrm{u}}^{+}$ transition, we obtain a value $k_B = 1.380(81) \cdot 10^{-23}$ J/K, which is in agreement, but less accurate than the literature value $k_B = 1.3806477(17) \cdot 10^{-23} $ J/K \cite{Codata}.

\section{CONCLUSION AND OUTLOOK}

In conclusion, we have investigated absorption and emission spectra of molecular dimers at high-pressure noble buffer gas conditions and demonstrated that they well follow the Boltzmann-type Kennard-Stepanov scaling. Lines of evidence for a thermal equilibrium distribution of vibrational states in the electronically excited state manifold are (i) observation of an emission spectrum being independent of the excitation for a broad parameter range, and the Kennard-Stepanov scaling of absorption and emission being well fulfilled for different molecular (ii) species, (iii) transitions, and (iv) temperatures. Further, the validity of the Kennard-Stepanov relation has been generalized to the use of nonlinear excitation techniques. Finally, accurate temperature values have been derived from this spectroscopic thermodynamic scaling, as possible in the well-controllable AMO-physics system.

For the future, we expect that ultrafast spectroscopy connecting molecular levels with different thermalization rates in the buffer gas can explore novel molecular wavepacket dynamic effects. An exciting further perspective is the development of new molecular laser sources based on collisional redistribution, transferring the progress made on alkali lasers \cite{Krupke} to molecular-noble buffer gas systems. Finally, accurate measurements of the Boltzmann-type frequency scaling  between absorption and emission in high-pressure buffer gas systems are a promising approach for future precise frequency-based determinations of both the local temperature and the Boltzmann constant.

We enthusiastically acknowledge discussions with F. Vewinger and G. Pichler. This work was supported by the DFG, grant number We1748-15.


\bibliography{draftapssamp}

\end{document}